\documentstyle[12pt,prd,aps,epsfig,preprint]{revtex}
\begin{document}
\tightenlines\draft
\date{\today}
\title{The $a_0K^+K^-$-vertex in light cone QCD sum rules }

\author{A. Gokalp~\thanks{agokalp@metu.edu.tr} and
        O. Yilmaz~\thanks{oyilmaz@metu.edu.tr}}
\address{ {\it Physics Department, Middle East Technical University,
06531 Ankara, Turkey}}

\maketitle

\begin{abstract}
We investigate the $a_0K^+K^-$-vertex in the framework of light
cone QCD sum rules. We estimate the coupling constant
g$_{a_0K^+K^-}$ which is an essential  ingredient in the analysis
of physical processes involving $a_0(980)$ meson. Our result is
somewhat  larger than the previous determinations of this coupling
constant.
\end{abstract}

\thispagestyle{empty} ~~~~\\
\pacs{PACS numbers:  12.38.Lg;13.25.Jx;14.40.Cs}

\newpage
\setcounter{page}{1}
%%%

The structure of light scalar mesons $f_0(980)$ and $a_0(980)$ has
been a controversial problem in hadron spectroscopy. In the naive
quark model $q\overline{q}$ \cite{R1}, $a_0(980)$ can be
interpreted as $a_0=(u\overline{u}-d\overline{d})/\sqrt{2}$. On
the other hand, the strong coupling of $f_0(980)$ to kaons
suggests the structure $f_0=s\overline{s}$ for this state.
However, then the almost exact degeneracy of the masses of
$a_0(980)$ and $f_0(980)$ cannot be explained. In order to explain
the properties of these states several proposals have been put
forward over the years. A four quark $q^2\overline{q}^2$ state
interpretation with symbolic quark structure
$f_0=s\overline{s}(u\overline{u}+d\overline{d})/\sqrt{2}$ and
$a_0=s\overline{s}(u\overline{u}-d\overline{d})/\sqrt{2}$ was
proposed in the framework of MIT-bag model where the scalar meson
states are spatially compact \cite{R2}. Another possibility about
the structure of $a_0(980)$ and $f_0(980)$ was suggested where
these meson states are considered to be bound states of hadrons.
This possibility is referred to as their being $K\overline{K}$
molecules in which case they are considered as spatially extended
objects \cite{R3}. Furthermore, some analyses suggest the
qualitative picture that these scalar meson states have a compact
$q^2\overline{q}^2$ structure that spends some part of its
lifetime in the $K\overline{K}$ meson system \cite{R4}.

The radiative decays of $\phi(1020)$ meson $\phi\rightarrow
a_0\gamma$ and $\phi\rightarrow f_0\gamma$ provide important tests
to distinguish among the different possibilities about the
structure of $a_0(980)$ and $f_0(980)$ scalar meson states
\cite{R4,R5}. It is generally agreed that the experimental data
supports the kaon loop mechanism for these decays in both the
$q^2\overline{q}^2$ state and $K\overline{K}$ molecule models
where these radiative decays proceed by photon emission from an
intermediate $K^+K^-$ loop \cite{R5}. Moreover, the study of the
reactions $\phi\rightarrow \pi^0\pi^0\gamma$ and $\phi\rightarrow
\pi^0\eta\gamma$ and of the interference patterns in these
reactions have been used to develop arguments about the structure
of $a_0(980)$ and $f_0(980)$ states \cite{R6}. In the analyses
involving $a_0(980)$ meson the strong coupling constant
g$_{a_0K^+K^-}$ plays an important role.

In this work, we estimate the coupling constant g$_{a_0K^+K^-}$ by
employing light cone QCD sum rules. This method has been applied
to study hadronic properties and in particular it has been used
for the calculation of hadronic coupling constants \cite{R7}.

In order to study the $a_0K^+K^-$-vertex and to estimate the
coupling constant g$_{a_0K^+K^-}$ we consider the two-point
correlation function
\begin{equation}\label{e1}
  T_{\mu}(p,q)=i\int d^{4}x e^{ip\cdot x}
  <K^+(q)|T\{j_\mu^K(x)j^{a_0}(0)\}|0>
\end{equation}
with $p$ and $j_{\mu}^K$ the four-momentum and the interpolating
current for $K^-$ meson, $j^{a_0}$ the interpolating current for
$a_0$ meson, and $q$ the four-momentum of $K^+$ state. The
interpolating quark currents are the axial vector
$j_{\mu}^K=\overline{u}\gamma_{\mu}\gamma_5s$ and the scalar
$j^{a_0}=(\overline{u} u-\overline{d} d)/2$ current. The scalar
current $j^{a_0}$ is assumed to have a non-vanishing matrix
element between the vacuum and $a_0(980)$ meson state, $<a_0\mid
j^{a_0}\mid 0>=\lambda_{a_0}$, where $\lambda_{a_0}$ is called the
overlap amplitude which was determined by QCD sum rules method
\cite{R8}, and this particular choice for the current in terms of
quark fields does not imply the pure $(\overline{u} u-\overline{d}
d)/\sqrt{2}$ structure for $a_0(980)$ meson.

The correlation function can be written in terms of two
independent invariant functions T$_1$ and T$_2$ as
\begin{equation}\label{e2}
  T_{\mu}(p,q)=iT_1\left((p+q)^2,~p^2\right)p_\mu+T_2\left((p+q)^2,~p^2\right)q_\mu~~.
\end{equation}
We consider the invariant function T$_1$. In order to construct
the theoretical part of the sum rule for the coupling constant
g$_{a_0K^+K^-}$ we calculate the function T$_1$ in terms of QCD
degrees of freedom by evaluating the correlation function in the
deep Euclidean region where $p^2$ and $(p+q)^2$ are large and
negative as an expansion near the light cone $x^2=0$. This
expansion involves matrix elements of non-local composite
operators between kaon and vacuum states which defines kaon
distribution amplitudes of increasing twist. We retain terms up to
twist four accuracy since higher twist amplitudes are known to
make a small contribution \cite{R9}. In our calculation we use the
full light propagator with both perturbative and nonperturbative
contributions which is given as \cite{R10}
\begin{eqnarray}\label{e3}
  iS(x,0)&=&<0|T\{\overline{q}(x)q(0)\}|0>\nonumber \\
         &=&i\frac{\not x}{2\pi^2x^4}-\frac{<\overline{q}q>}{12}-
         \frac{x^2}{192}m_0^2<\overline{q}q>\nonumber \\
         &~&-ig_s\frac{1}{16\pi^2}\int_0^1du\left\{
         \frac{\not{x}}{x^2}\sigma_{\mu\nu}G^{\mu\nu}(ux)
         -4iu\frac{x_\mu}{x^2}G^{\mu\nu}(ux)\gamma_\nu\right\}+...~~.
\end{eqnarray}
After a straightforward computation and performing the Fourier
transforms we obtain
\begin{eqnarray}\label{e4}
T_1\left(p^2,~(p+q)^2\right)&=&\frac{f_K M_K^2}{2(6m_s)}  \int_0^1
du \varphi_{\sigma K}(u)\frac{2[p+(1-u)q]\cdot
q}{\left\{[p+(1-u)q]^2\right\}^2} \nonumber \\
&&+\frac{1}{2}\frac{f_K M_K^2}{m_s}\int_0^1 du \varphi_{p
K}(u)\frac{1}{[p+(1-u)q]^2} \nonumber \\
&&+\frac{f_{3K}}{2}\int_0^1 dv \int D\alpha_i \varphi_{3
K}(\alpha_i)\frac{M_K^2}{\left\{[p+(1-\alpha_1-v\alpha_3)q]^2\right\}^2}(2v-1)~~.
\end{eqnarray}
In this expression the functions $\varphi_{\sigma K}$ and
$\varphi_{pK}$ are the twist 3 quark-antiquark kaon distribution
amplitudes defined by the matrix elements \cite{R11}
\begin{eqnarray}\label{e5}
<K(q)|\overline{u}(x)i\gamma_5 s(0)|0> =f_K \mu_K \int_0^1 du
e^{iuq\cdot x} \varphi_{pK}(u)~~,
\end{eqnarray}
\begin{eqnarray}\label{e6}
<K(q)|\overline{u}(x)\sigma_{\mu\nu}\gamma_5 s(0)|0> = i\frac{f_K
\mu_K}{6}\left(1-\frac{M_K^2}{\mu_K^2}\right)(q_\mu x_\nu-x_\mu
q_\nu) \int_0^1 du e^{iuq\cdot x} \varphi_{\sigma K}(u)~~,
\end{eqnarray}
where $\mu_K=M_K^2/m_s$ is the twist 3 distribution amplitude
normalization factor and we put $m_u=m_d=0$. We work in the gauge
$x^\mu A_\mu=0$, consequently the path-ordered gauge factor is not
included in the matrix elements. The twist 3 quark-antiquark-gluon
kaon distribution amplitude $\varphi_{3K}$ is defined as
\cite{R11}
\begin{eqnarray}\label{e7}
<K(q)|\overline{u}(x)\sigma_{\alpha\beta}g_s G{\mu\nu}(vx) s(0)|0>
&=& if_{3K}\left[(q_\mu q_\alpha g_{\nu\beta}-q_\nu q_\alpha
g_{\mu\beta})-(q_\mu q_\beta g_{\nu\alpha}-q_\nu q_\beta
g_{\mu\alpha})\right] \nonumber \\
&\times&\int D\alpha_i \varphi_{3 K}(\alpha_i) e^{iuq\cdot
x(\alpha_1+v\alpha_3)}
\end{eqnarray}
where $D\alpha_i=d\alpha_1 d\alpha_2 d\alpha_3
\delta(1-\alpha_1-\alpha_2-\alpha_3)$. After performing the Borel
transformation with respect to  the variables $Q_1^2=-(p+q)^2$ and
$Q_2^2=-p^2$, we obtain the theoretical expression for the
invariant function in the form
\begin{eqnarray}\label{e8}
T_1(M_1^2,~M_2^2)&=&\frac{f_KM_K^2 M^2}{2m_s} [- \varphi_{p
K}(u_0)+\frac{1}{6}\varphi_{\sigma K}^{\prime}(u_0)]
\nonumber \\
&+&
\frac{f_{3K}M_K^2}{2}\int_0^{u_0}d\alpha_1\int_{u_0-\alpha_1}^{1-\alpha_1}\frac{d\alpha_3}{\alpha_3}
\varphi_{3K}(\alpha_1,1-\alpha_1-\alpha_3,\alpha_3)\left(2\frac{u_0-\alpha_1}{\alpha_3}-1\right)
\end{eqnarray}
where M$_1^2$ and M$_2^2$ are the Borel parameters and
\begin{eqnarray}
  u_0=\frac{M_1^2}{M_1^2+M_2^2}~~~~,~~
  M^2=\frac{M_1^2M_2^2}{M_1^2+M_2^2} ~~. \nonumber
\end{eqnarray}
We like to note that if we multiply the correlation function by
the four-momentum p, we obtain
\begin{eqnarray}\label{e81}
  p^\mu T_{\mu}(p,q)=-\int d^{4}x e^{ip\cdot x}\left[
  <K^+(q)|T\{\frac{\partial}{\partial
  x_\mu}j_\mu^K(x)j^{a_0}(0)\}|0>\right.\nonumber \\
\left.+\delta(x_0)<K^+(q)|[j_\nu^K(x),~j^{a_0}(0)]|0>\right]~,
\end{eqnarray}
where the second term results from the differentiation of the
function $\theta(x_0)$ in the T-product of the currents. In the
SU(3)$_{fl}$ limit $\partial^\mu j_\mu^K(x)=0$, thus the first
term on the right hand side of Eq. (9) vanishes. The second term
can be calculated using the standard commutation relations,
yielding for the correlation function the Ward identity
\begin{eqnarray}\label{e82}
  p^\mu T_{\mu}(p,q)=-if_Kq_\nu. \nonumber
\end{eqnarray}
Similar Ward identities were considered in \cite{R11} where they
were used to obtain relations between various pion distribution
amplitudes.

Two-point correlation function satisfies a dispersion relation,
therefore we can represent the invariant function as
\begin{equation}\label{e9}
  T_1\left((p+q)^2,~p^2\right)=\int\int ds ds'~
  \frac{\rho^{had}(s,s')}{[s-(p+q)^2)](s'-p^2)}~~.
\end{equation}
We saturate this dispersion relation by inserting a complete set
of one hadron-states into the correlation function and we consider
the single-particle K and $a_0$ states, this way we obtain
\begin{eqnarray}\label{e10}
T_1\left((p+q)^2,~p^2\right)=&&\frac{<0\mid j_{\mu}^K\mid
K(p)><KK\mid a_0> <a_0(p+q)|j^{a_0}\mid 0>}
  {\left[(p+q)^2-M_{a_0}^2\right]\left(p^2-M_K^2\right)} \nonumber \\
  &&+\int_{s_{0}} ds\int_{s'_{0}}ds'~\frac{\rho^{cont}(s,s')}{[s-(p+q)^2](s'-p^2)}~~,
\end{eqnarray}
where the hadronic spectral density includes the contributions of
higher resonances and the hadronic continuum. The matrix element
$<KK\mid a_0>$ defines the coupling constant g$_{a_0K^+K^-}$
\begin{eqnarray}\label{e11}
<K^+(q)K^-(p)\mid a_0(p+q)>=g_{a_0K^+K^-}
\end{eqnarray}
and the current-particle matrix elements are given as
\begin{eqnarray}\label{e12}
<a_0(p+q)\mid j^{a_0}\mid 0>=\lambda_{a_0}~~,
\end{eqnarray}
\begin{eqnarray}\label{e13}
<0\mid j_{\mu}^K\mid K(p)>=if_Kp_\mu~~.
\end{eqnarray}
After performing a similar double Borel transformation we obtain
for the hadronic representation the result
\begin{eqnarray}\label{e14}
T_1(M_1^2,M_2^2)&=&\lambda_{a_0}f_Kg_{a_0K^+K^-}e^{-M^2_{a_0}/M_1^2}e^{-M^2_K/M_2^2}
\nonumber \\
&+&\int_{s_{0}}ds\int_{s'_{0}}ds'~\rho^{cont}(s,s')e^{-s/M_1^2}e^{-s'/M_2^2}~~.
\end{eqnarray}
The sum rule for the coupling constant g$_{a_0K^+K^-}$ then
follows by equating the expressions $T_1(M_1^2,M_2^2)$ obtained
for the invariant function $T_1((p+q)^2,p^2)$ by theoretical (Eq.
8) and by physical (Eq. 15) considerations. In order to do this we
have to identify the second term in Eq. 15 representing the
continuum contribution with a part of the term calculated
theoretically in QCD, thus affecting the subtraction of the
continuum. The prescription  that has been suggested for this
purpose \cite{R12,R13} is based on the observation that the
distribution amplitudes $\varphi_{pK}(u)$ and $\varphi_{\sigma
K}(u)$ are polynomials in $(1-u)$, therefore we can write
\begin{eqnarray}\label{e15}
-\varphi_{p}(u)+\frac{1}{6}\varphi^\prime_{\sigma}(u)=\sum_{k=0}^{N}b_k(1-u)^k~~.
\end{eqnarray}
The continuum subtraction is affected in the leading twist 3
quark-antiquark term, since the contribution of the twist 3
quark-antiquark-gluon term in Eq. 8 is small, therefore finally we
obtain the sum rule for the coupling constant g$_{a_0K^+K^-}$ in
the form
\begin{eqnarray}\label{e16}
g_{a_0K^+K^-}&=&\frac{1}{2\lambda_{a_0}}e^{M^2_{a_0}/M_1^2}e^{M^2_K/M_2^2}
\left\{
\frac{M^2M_K^2}{m_s}\sum_{k=0}^{N}b_k\left(\frac{M^2}{M_1^2}\right)^k
\left[1-e^{-A}\sum_{i=0}^{k}\frac{A^i}{i!}+e^{-A}\frac{M^2M_K^2}{M_1^2M_2^2}\frac{A^{(k+1)}}{(k+1)!}\right]
\right. \nonumber \\
&+&\left.
\frac{f_{3K}M_K^2}{f_K}\int_0^{u_0}d\alpha_1\int_{u_0-\alpha_1}^{1-\alpha_1}\frac{d\alpha_3}{\alpha_3}
\varphi_{3K}(\alpha_1,1-\alpha_1-\alpha_3,\alpha_3)\left(2\frac{u_0-\alpha_1}{\alpha_3}-1\right)\right\}
\end{eqnarray}
where $A=s_0/M^2$ with $s_0$ the smallest continuum threshold.

In the numerical evaluation of the sum rule we use the twist 3
kaon distribution amplitudes given by \cite{R11}
\begin{eqnarray}\label{e17}
\varphi_{pK}(u)=1&+&\left(30\frac{f_{3K}}{\mu_Kf_K}-\frac{5}{2}\frac{M_K^2}{\mu_K^2}\right)C_2^{1/2}(2u-1)
\nonumber \\
&+&\left[-3\frac{f_{3K}\omega_{3K}}{\mu_Kf_K}-\frac{27}{20}\frac{M_K^2}{\mu_K^2}(1+6a_2^K)\right]C_4^{3/2}(2u-1)
\end{eqnarray}
\begin{eqnarray}\label{e18}
\varphi_{\sigma K}(u)=
6u\overline{u}\left\{1+\left[5\frac{f_{3K}}{\mu_Kf_K}(1-\frac{1}{10}\omega_{3K})
-\frac{7}{20}\frac{M_K^2}{\mu_K^2}(1+\frac{12}{7}a_2^K)\right]C_2^{3/2}(2u-1)\right\}
\end{eqnarray}
\begin{eqnarray}\label{e19}
\varphi_{3K}(u)=360\alpha_1\alpha_2\alpha_3^2\left[1+\frac{\omega_{3K}}{2}(7\alpha_3-3)\right]
\end{eqnarray}
where $C_m^k(2u-1)$ are the Gegenbauer polynomials. The overlap
amplitude $\lambda_{a_0}$ has been determined previously as
$\lambda_{a_0}=(0.21\pm 0.05)~~GeV^2$ employing QCD sum rules
method \cite{R8}. We also adopt the values at the renormalization
scale 1 GeV $m_s(1GeV)=150~~MeV$, in SU(3)$_{fl}$ limit
$f_{3K}(1GeV)=f_{3\pi}(1GeV)=0.0035~~GeV^2$,
$\omega_{3K}(1GeV)=-2.88$ and $f_K=0.160~~GeV$ \cite{R11} with
$M_K=0.4937~~MeV$.

We then study the dependence of the sum rule for the coupling
constant g$_{a_0K^+K^-}$ on the continuum threshold $s_0$ and on
the Borel parameters $M_1^2$ and $M_2^2$ by considering
independent variations of these parameters. We find that the sum
rule is quite stable for the range of these parameters $1.00\leq
s_0\leq 1.10~~GeV^2$, $0.7\leq M_1^2\leq 1.4~~GeV^2$ and $2\leq
M_2^2\leq 6.0~~GeV^2$. By varying the values of the parameters
$s_0$, $M_1^2$, and $M_2^2$ in these regions  we obtain the result
for the coupling constant g$_{a_0K^+K^-}$ as $4.4\leq
g_{a_0K^+K^-}\leq 5.6~~GeV$. The variation of the coupling
constant as a function of the Borel parameters $M_1^2$ and
$M_2^2$, and the continuum threshold $s_0$ is shown in Fig. 1. We
note that the sign that we obtain for the coupling constant
g$_{a_0K^+K^-}$ is negative, that is g$_{a_0K^+K^-}<0$.

There has been several previous estimations of the coupling
constant g$_{a_0K^+K^-}$. The Novosibirsk SND collaboration data
of the radiative decay $\phi\rightarrow \pi^0\eta\gamma$
\cite{R14} was analyzed in a phenomenological framework in which
the contributions of $\rho$ meson, chiral loop and $a_0$ meson
were considered, and the value g$_{a_0K^+K^-}=(-1.5\pm 0.3)~~GeV$
was obtained for this coupling constant \cite{R15}. The coupling
constant thus obtained results in constructive interference
between the contribution of different amplitudes. From the
analysis of their experimental data of $\phi\rightarrow
\pi^0\eta\gamma$ decay, the KLOE collaboration estimated the
coupling constant g$_{a_0K^+K^-}$ as g$_{a_0K^+K^-}=(2.3\pm
0.7)~~GeV$ \cite{R16}. A new analysis of the KLOE data on
$\phi\rightarrow \pi^0\eta\gamma$ decay, on the other hand, gives
the result g$_{a_0K^+K^-}=(2.63^{+1.84}_{-1.28})~~GeV$ \cite{R17}.
In this analysis the phase $\delta$ of the interference between
$\phi\rightarrow a_0\gamma\rightarrow \pi^0\eta\gamma$ and
$\phi\rightarrow \rho^0\pi^0\rightarrow \pi^0\eta\gamma$
amplitudes was obtained as $\delta=0$, which is in accordance with
the constructive interference of the different amplitudes observed
in the phenomenological analysis of $\phi\rightarrow
\pi^0\eta\gamma$ decay \cite{R14}. Our estimation of the coupling
constant g$_{a_0K^+K^-}$ using light cone QCD sum rules results in
a  value somewhat larger than the previous determinations based on
the analysis of $\phi\rightarrow \pi^0\eta\gamma$ data. However,
because of the intrinsic uncertainties of the light cone QCD sum
rule method, our result can be taken only to indicate that the
scalar $a_0$ meson state may have somewhat large strong coupling.
Finally we note that the sign of the coupling constant
g$_{f_0K^+K^-}$  was obtained as g$_{f_0K^+K^-}>0$ in a previous
light cone QCD sum rule determination of this coupling constant
\cite{R13}, and in a phenomenological analysis of $\phi\rightarrow
\pi^0\pi^0\gamma$ decay \cite{R18}. Thus our result about
g$_{a_0K^+K^-}$ that is g$_{a_0K^+K^-}<0$, is consistent with the
result obtained in the $q^2\overline{q}^2$ model where
g$_{f_0K^+K^-}=-g_{a_0K^+K^-}$ \cite{R2,R5,R19}.

It has been argued that the ratio $R=BR(\phi\rightarrow
f_0\gamma)/BR(\phi\rightarrow a_0\gamma)$ can provide insight into
the structure of the scalar $a_0(980)$ and $f_0(980)$ mesons
\cite{R4}. The KLOE collaboration obtained this ratio as
\cite{R16}
\begin{eqnarray}\label{e20}
R=\frac{BR(\phi\rightarrow f_0\gamma)}{BR(\phi\rightarrow
a_0\gamma)}=6.1\pm 0.6~~.
\end{eqnarray}
This ratio can be written, assuming the intermediate
$K\overline{K}$ loop mechanism for these decays, in the form
\cite{R4}
\begin{eqnarray}\label{e21}
R=\frac{g^2_{f_0K^+K^-}}{g^2_{a_0K^+K^-}}~\frac{F^2_{f_0}(R)}{F^2_{a_0}(R)}~cot^2\theta
\end{eqnarray}
where the factors $F^2_{f_0}(R)$ and $F^2_{a_0}(R)$ are related to
the spatial extensions of $f_0(980)$ and $a_0(980)$ mesons, and
for point-like effective field theory calculations $F^2(R)=1$. For
a spatially extended system with r.m.s. radius
$R>O(\Lambda^{-1}_{QCD})$ the high momentum region of the
integration is suppressed \cite{R4}, resulting in the form factor
with the property $F^2(R)<1$, conversely $F^2(R)\rightarrow 1$
means a spatially compact system. The angle $\theta$ is the
isospin mixing angle in the $f_0-a_0$ system. If we use the result
$6.2\leq g_{f_0K^+K^-}\leq 7.8~~GeV$ obtained by a light cone QCD
sum rule calculation \cite{R13}, and our result $4.4<\mid
g_{f_0K^2K^-}\mid < 5.6~~GeV$ obtained by a similar light cone QCD
sum rule method calculation we obtain
\begin{eqnarray}\label{e22}
cot\theta~\frac{F_{f_0}(R)}{F_{a_0}(R)}\sim 1.8 ~~. \nonumber
\end{eqnarray}
Since in light cone QCD sum rule calculations isospin is assumed
to be exact, which corresponds to $\theta=45^o$, we thus find
$F_{f_0}(R)/F_{a_0}(R)\sim 1.8$ which seems to imply that the
spatial extensions of $f_0(980)$ and $a_0(980)$ mesons are not
equal. Therefore, also given the relation that we obtain about the
relative sign between the coupling constants g$_{f_0K^+K^-}$ and
g$_{a_0K^+K^-}$ in accordance with $q^2\overline{q}^2$ model, we
may suggest that our result supports the view that the structure
of $a_0(980)$ and  $f_0(980)$ mesons is a combination of a
$K\overline{K}$ molecule with a compact $q^2\overline{q}^2$ core
with the spatial extension of $f_0(980)$ being more compact than
that of $a_0(980)$ meson.

%\pagebreak

\newpage

\begin{figure}\hspace{-0.5cm}
\epsfig{figure=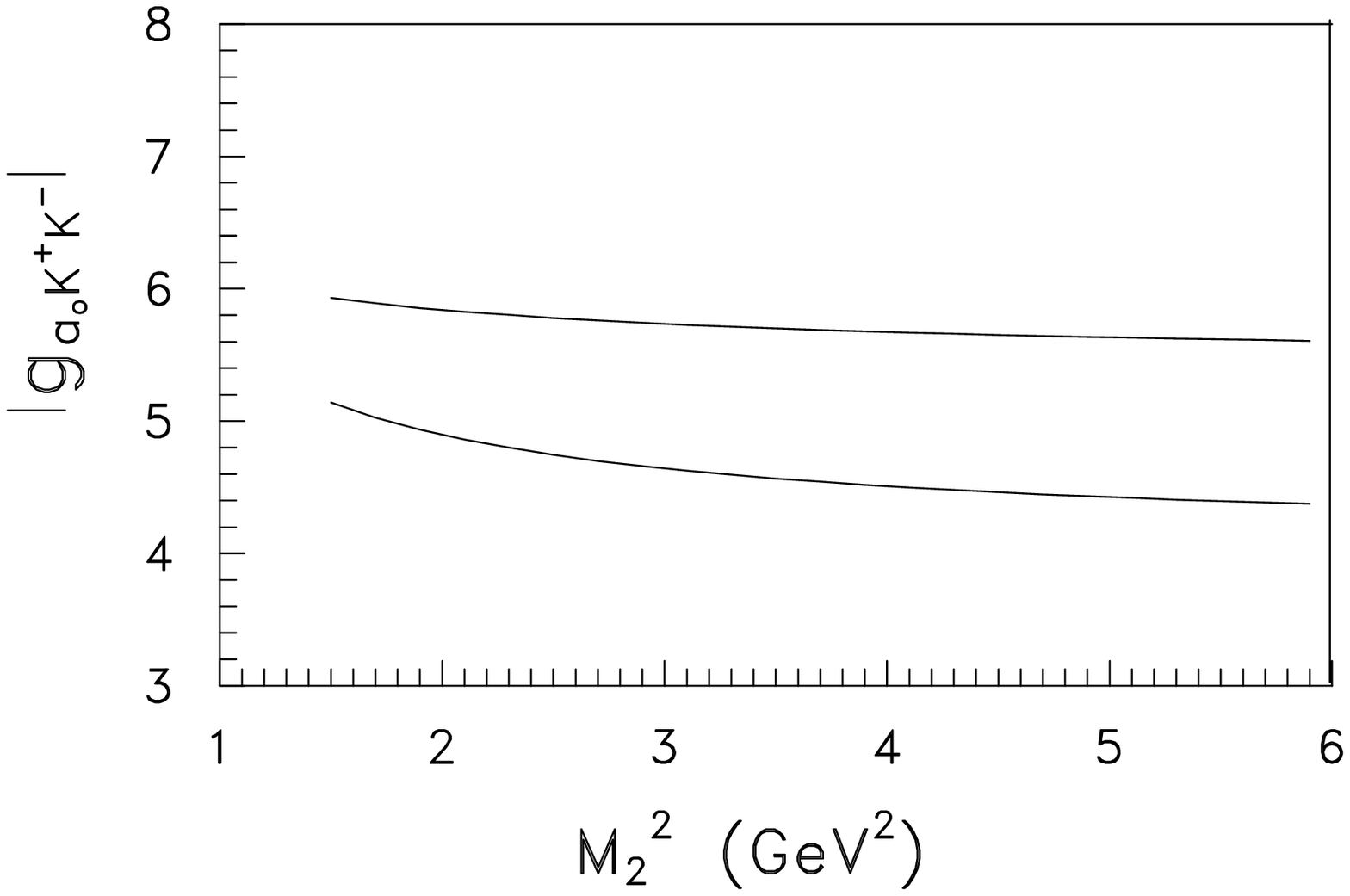,width=12cm,height=18cm} \vspace*{-3.5cm}
\caption{The coupling constant g$_{a_0K^+K^-}$ as a function of
the Borel parameter $M_2^2$ for different values of the threshold
parameter $s_0$ and the Borel parameter $M_1^2$. The curves denote
the limits of the stability region.} \label{fig1}
\end{figure}

\end{document}